\documentclass[%
reprint,
 amsmath,amssymb,
 aps,
]{revtex4-2}

\usepackage{graphicx}% Include figure files
\usepackage{dcolumn}% Align table columns on decimal point
\usepackage{bm}% bold math
\usepackage[dvipsnames]{xcolor}
\usepackage{tabu}
\usepackage[margin=1in]{geometry}
\usepackage{hyperref}% add hypertext capabilities
\usepackage{outlines}
\usepackage{multirow}
\usepackage{url}
\newcommand{\Cro}{Cr$_2$O$_3$ }
\newcommand{\Feo}{Fe$_2$O$_3$ }
\newcommand{\aFeo}{$\alpha$-Fe$_2$O$_3$ }
\newcommand{\Crop}{Cr$_2$O$_3$}
\newcommand{\Feop}{Fe$_2$O$_3$}
\newcommand{\aFeop}{$\alpha$-Fe$_2$O$_3$}

\newcommand{\mc}{$m_{\textrm{\tiny{Cr}}}$}
\newcommand{\ac}{$a_{\textrm{\tiny{Cr}}}$}
\newcommand{\tc}{$t_{\textrm{\tiny{Cr}}}$}
\newcommand{\qc}{$q_{\textrm{\tiny{Cr}}}$}

\newcommand{\mf}{$m_{\textrm{\tiny{Fe}}}$}
\newcommand{\af}{$a_{\textrm{\tiny{Fe}}}$}
\newcommand{\tf}{$t_{\textrm{\tiny{Fe}}}$}
\newcommand{\qf}{$q_{\textrm{\tiny{Fe}}}$}
\begin{document}

\preprint{APS/123-QED}

\title{Hidden orders and (anti-)Magnetoelectric Effects in \Cro and \aFeop}

\author{Xanthe H. Verbeek}
 \email{xverbeek@ethz.ch}
\author{Andrea Urru}
\author{Nicola A. Spaldin}
\affiliation{Materials Department, ETH Zürich,\\ 8093 Zürich, Switzerland
}

\date{\today}% It is always \today, today, but any date may be explicitly specified

\begin{abstract}
We present \textit{ab initio} calculations of hidden magnetoelectric multipolar order in \Cro and its iron-based analogue, \aFeop. First, we discuss the connection between the order of such hidden multipoles and the linear magnetoelectric effect. Next, we show the presence of hidden antiferroically-ordered magnetoelectric multipoles in both the prototypical magnetoelectric material \Crop, and centrosymmetric \aFeop, which has the same crystal structure as \Crop, but a different magnetic dipolar ordering. In turn, we predict anti-magnetoelectric effects, in which local magnetic dipole moments are induced in opposite directions under the application of an external electric field, to create an additional antiferromagnetic ordering. We confirm the predicted induced moments using first-principles calculations. Our results demonstrate the existence of hidden magnetoelectric multipoles leading to local linear magnetoelectric responses even in centrosymmetric materials, where a net bulk linear magnetoelectric effect is forbidden by symmetry.
\end{abstract}

\maketitle

%\section{\label{sec:intro}Introduction}
Magnetoelectric (ME) materials, show a net change in magnetization $M$ when an external electric field $\mathcal{E}$ is applied or, vice-versa, change their electric polarization $P$ in the presence of a magnetic field $\mathcal{H}$ \cite{dzyaloshinskyMagnetoelectricalEffectAntiferromagnets1960}. 
These materials have been a subject of active research \cite{halleySizeinducedEnhancedMagnetoelectric2014,spaldinRenaissanceMagnetoelectricMultiferroics2005,shiratsuchiMagnetoelectricControlAntiferromagnetic2018}, as the coupling of magnetic and electric degrees of freedom is potentially useful, with proposed applications including low-energy-consumption memory devices, sensors, and transistors \cite{kotaEnhancementSpinCorrelation2014,fiebigRevivalMagnetoelectricEffect2005}. 
The lowest-order, \textit{linear}, contribution to the ME response \cite{Landau_electrodynamics}, which requires the simultaneous breaking of space- and time-inversion symmetries, was established to be linked to a microscopic order of ME multipoles \cite{Zeldovich:1958,Gorbatsevich/Kopaev/Tugushev:1983,Chatel/Buin:2002, Ederer_Spaldin_toroidal_moment, spaldinMonopolebasedFormalismDiagonal2013}, which are odd-parity, second-order multipoles of the magnetization density $\boldsymbol{\mu}(\boldsymbol{r})$. In their irreducible spherical form, the ME multipoles are the scalar ME monopole ($a$), the ME toroidal moment vector ($\boldsymbol{t}$) and the ME quadrupole tensor ($q$), 
\begin{align}
\label{eq:ME_mult}
    a = & \frac{1}{3} \int \boldsymbol{r} \cdot \boldsymbol{\mu}(\boldsymbol{r}) d^3\boldsymbol{r}, \\
    t_{i} = & \frac{1}{2} \int [ \boldsymbol{r} \times \boldsymbol{\mu}(\boldsymbol{r})]_{i} \thickspace d^3\boldsymbol{r}, \\
    q_{ij} = & \frac{1}{2} \int \big[ r_i\mu_j(\boldsymbol{r}) +r_j\mu_i(\boldsymbol{r}) -\frac{2}{3}\delta_{ij}\boldsymbol{r} \cdot \boldsymbol{\mu}(\boldsymbol{r})\big] d^3\boldsymbol{r},
\end{align}
which correspond respectively to the trace, the anti-symmetric part and the symmetric traceless part of the ME multipole tensor \cite{spaldinMonopolebasedFormalismDiagonal2013}, defined as 
\begin{align}
    \mathcal{M}_{ij} = & \int r_i\mu_j(\boldsymbol{r})d^3\boldsymbol{r}. \label{eq:mult_tensor}
\end{align}
ME multipoles provide a handle for understanding and predicting the linear ME effect starting from the microscopic environment, since they have a one-to-one link to the linear ME tensor $\alpha_{ij}$, defined as $\alpha_{ij} =\mu_0 \partial M_j / \partial \mathcal{E}_i \rvert_{\mathcal{H}}$, with $\mu_0$ the vacuum permeability. Specifically, monopoles $a$ and $q_{x^2-y^2}$, $q_{z^2}$ quadrupoles account for the diagonal isotropic and anisotropic linear ME effect, whereas the toroidal moments $t_{i}$ and the $q_{xy}$, $q_{xz}$, and $q_{yz}$ quadrupoles are linked to the off-diagonal anti-symmetric and symmetric linear ME effect, respectively. Analogously, the second-order ME effect can be captured by the next-higher order magnetic multipoles, the magnetic octupoles \cite{urru_magnetic_2022}.
$\mathcal{M}_{ij}$ can be split into a local atomic-site contribution, which we will focus on in the following, and an origin-dependent, multivalued contribution, written in terms of products of the atomic positions and magnetic moments \cite{spaldinMonopolebasedFormalismDiagonal2013}. In linear ME materials, the simultaneous breaking of time- and space-inversion symmetries allows for a ferroic order of local ME multipoles, consistent with a net bulk ME effect. The set of allowed ferroically-ordered ME multipoles is determined by the magnetic point group symmetry. Antiferroically-ordered local ME multipoles can also be symmetry allowed \cite{Thole_thesis, spaldinMonopolebasedFormalismDiagonal2013}, although they do not give any net contribution to the ME tensor. In centrosymmetric magnetic materials, however, they should in principle provide the lowest order, \emph{local} ME response, even though a net ME effect is forbidden by symmetry.

\begin{figure}[t]
 \centering
 \includegraphics[trim={6.8cm 2.8cm 11.1cm 2.2cm},clip, width=0.45\textwidth]{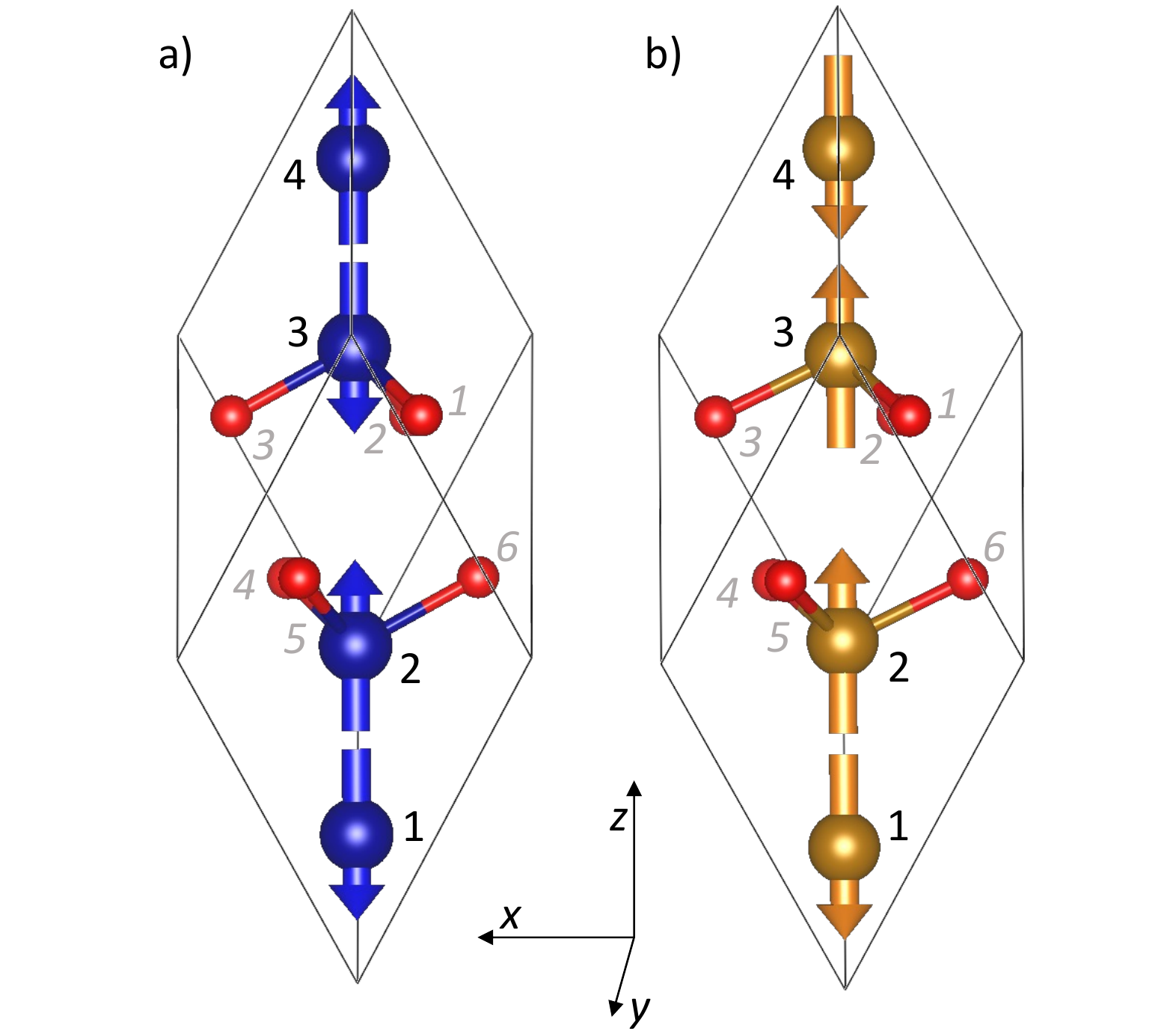}
\caption{Crystal structure and magnetic order of a) \Crop, b) \Feop. Cr, Fe and O ions are colored blue, gold and red, respectively. Magnetic moments are indicated by arrows. The atoms are numbered following the conventional order of the Wyckoff positions. The magnetic easy axis is parallel to the $z$-axis.}
\label{fig:Mag_compare}
\end{figure}

In this work, we analyze the link between the multipolar order and the local, atomic ME response in the isostructural materials \Cro and \aFeo (from now on \Feop). Both materials adopt the corundum structure, with the centrosymmetric point group $\bar{3}$m (space group R$\bar{3}$c), and are easy-axis antiferromagnets, below 307 K \cite{foner_high-field_1963, muEffectSubstitutionalDoping2013} and 263 K \cite{morin_electrical_1951, dzyaloshinskyThermodynamicTheoryWeak1958, Fe2O3_footnote}, respectively. Importantly, however, they have different magnetic orderings, as shown in Fig. \ref{fig:Mag_compare}. Specifically, the magnetic order in \Cro breaks both inversion and time-reversal symmetries, whereas in \Feo it breaks time-reversal symmetry only. As a result, \Cro is a well known ME material \cite{hornreichStatisticalMechanicsOrigin1967, brownStudyMagnetoelectricDomain1998, lataczMagnetoelectricEffectCr2O32000, ye_dynamical_2014, kosub_purely_2017}, in which the linear ME effect was proposed \cite{dzyaloshinskyMagnetoelectricalEffectAntiferromagnets1960} and measured \cite{astrovMagnetoelectricEffectChromium1961} for the first time, whereas \Feo does not show a net linear ME effect, and instead its symmetry allows a non-relativistic, altermagnetic spin splitting \cite{smejkal_altermagnetism_2022, yuan_prediction_2021}. Despite the difference in global symmetry, the local site symmetries are similar in \Cro and \Feop. Local atomic ME multipoles and, in turn, a local ME response, are allowed in both compounds, as they only require breaking of the $local$ inversion symmetry, which happens at the specific Wyckoff sites occupied by the Cr, Fe, and O atoms in both materials. Thus, we expect that locally some of the special physics seen in \Cro may be preserved in \Feop. In this work we explore this possibility. Our main finding is that \Feo indeed has a hidden antiferromultipolar order that leads to a local anti-ME response, with a strength that is comparable with that in ME \Crop. \\
%
%\section{\label{sec:method}Methods}

Our density-functional calculations were performed within the non-collinear local spin density approximation (LSDA+U) \cite{LDA_pz}, with spin-orbit interaction included, as implemented in the plane-wave code VASP \cite{VASP_1, VASP_2} and in the full-potential linearized augmented-plane wave (FP-LAPW) code \texttt{ELK} \cite{ELK}. Correlation effects were dealt with by applying the rotationally invariant Hubbard U correction \cite{Dudarev_U} on Fe (Cr) $d$ states, with U $=5.5$ ($4.0$) eV and J $=0.5$ ($0.5$) eV. 
Structural relaxations and force-constant calculations were performed in VASP, using projector-augmented wave (PAW) pseudopotentials \cite{blochl_projector_1994} (valence electrons: Cr 3p$^6$3d$^5$4s$^1$, Fe 3d$^7$4s$^1$, O 2s$^2$2p$^4$, PAW datasets Cr\_sv, Fe\_sv, O) with a kinetic energy cut-off of 800 eV for the wavefunctions. Brillouin Zone (BZ) integrations were performed using a uniform $\Gamma$-centered $7\times7\times7$ Monkhorst-Pack k-point mesh \cite{Monkhorst_Pack_mesh}. With these parameters we have accurately captured known experimental band gaps and magnitudes of the magnetic moments. We found equilibrium lattice constants $a = 5.31 \, \mathrm{\AA}, \thickspace \alpha = 54.87^{\circ} $ for \Crop, 0.78\% and 0.26\% smaller than experiment, respectively \cite{hill_crystallographic_2010}. For \Feop, we found $a = 5.35 \, \mathrm{\AA}, \thickspace \alpha = 55.25^{\circ} $, 1.44\% smaller and 0.03\% larger than experiment, respectively \cite{hillNeutronDiffractionStudy2008, lattice_parameter_footnote}. 
The spin contributions to the local diagonal and off-diagonal lattice-mediated ME response in the xy-plane were obtained by calculating the magnetic moments induced by an applied electric field. This response is derived from an appropriate superposition of the magnetic moments induced by freezing in the eigendisplacements of the four twofold degenerate $E_u$ infrared-active phonon modes \cite{iniguezFirstPrinciplesApproachLatticeMediated2008}. Note that in contrast to Ref. \cite{iniguezFirstPrinciplesApproachLatticeMediated2008}, we focus on the \emph{local atomic} magnetic response.
The local magnetic multipoles and ME responses were calculated using the \texttt{ELK} code. The Cr, Fe, and O ion cores were described using muffin-tin spheres with radii 1.0716 \AA, 1.0400 \AA \space and 0.80435 \AA, respectively; the APW functions and the potential were expanded in a spherical harmonics basis, with cut-offs $l_{\text{max} (\text{apw})} = l_{\text{max} (\text{o})} = 12$. The BZ was sampled using a $6 \times 6 \times 6$ $\Gamma$-centered k-point mesh. 

When freezing in the phonons, the resulting changes in the local magnetic moments are small at relevant phonon amplitudes, thus extensive convergence tests with respect to a wide range of variables were performed (see the discussion in the Supplementary Material). The angular parts of the multipoles were calculated by decomposing the density matrix into its irreducible spherical tensors and extracting the relevant components \cite{multipole_decomposition}.

%\section{\label{sec:results}Results \lowercase{and} Discussion}
%\subsection{\label{sec:mult_sym} Multipole and Symmetry analysis}
As described above, the linear ME effect requires time-reversal and inversion symmetries to be broken. This is the case in \Crop, but in \Feo the global inversion symmetry is preserved. We determine which multipoles are allowed and their subsequent arrangement by studying both how each multipole transforms and how the atoms permute under the 12 symmetry operations of the R$\bar{3}$c space group (for more details on the symmetry analysis see the Supplementary Material). 
We find that on the transition metal (TM) ions in \Cro and \Feop, whose Wyckoff site symmetry (3) is the same, ME monopoles $a$, $t_z$ toroidal moments and $q_{z^{2}}$ quadrupoles are allowed, but with different ordering. We support the results of our symmetry analysis by first-principles calculations of the multipole components in \Cro and \Feo at their respective equilibrium structures. These calculations confirm the multipolar ordering obtained from the symmetry analysis and give us the absolute (as opposed to relative) sign of the multipoles, reported in Table \ref{tab:Multi_Cro_Feo_TM}, for the antiferromagnetic domains shown in Fig. \ref{fig:Mag_compare}. Furthermore, we calculate the size of the magnetic dipoles ($2.6 \, \mu_{\text{B}}$ and $4.1 \, \mu_{\text{B}}$ for Cr and Fe, respectively) and the angular parts of $a$, $t_z$, and $q_{z^{2}}$ ( $3\times10^{-3}$, $2\times10^{-5}$, and $2\times10^{-3}$ $\mu_{\text{B}}$ in \Cro and $4\times10^{-3}$, $7\times10^{-5}$, and $4\times10^{-3}$ $\mu_{\text{B}}$ in \Feop, respectively). We note that the multipole magnitudes are of similar size in both materials, approximately scaling with the size of the dipole moments. 
\begin{table}[t]
    \centering
    {\renewcommand{\arraystretch}{1.3}
    \begin{tabular}{l|c|c|c|c|c|c|c|c|c}
        \hline
        & \multicolumn{4}{c|}{\Cro} & & \multicolumn{4}{c}{\Feo} \\
        \cline{2-10}
        & Cr$_1$ & Cr$_2$ & Cr$_3$  & Cr$_4$ & &Fe$_1$ & Fe$_2$ & Fe$_3$  & Fe$_4$\\
        \hline
        $m_z$& $-$\mc & \mc & $-$\mc & \mc & & $-$\mf & \mf & \mf &  $-$\mf \\
        $a$  &  $-$\ac &  $-$\ac &  $-$\ac &  $-$\ac & &  \af &  \af  &   $-$\af &  $-$\af \\
        $t_z$ & \tc &  $-$\tc  &  $-$\tc &  \tc & &  \tf &  $-$\tf &  \tf &  $-$\tf \\
        $q_{z^{2}}$ & $-$\qc & $-$\qc &  $-$\qc &  $-$\qc & & \qf & \qf &  $-$\qf &  $-$\qf \\
         \hline
        \end{tabular}     
   }
    \caption{Symmetry-allowed magnetic moments and ME multipoles, and their ordering on the transition metal ions in \Cro and \Feop. Atoms are labelled as in Fig. \ref{fig:Mag_compare}.}
    \label{tab:Multi_Cro_Feo_TM}
\end{table}
The ferroic ordering of $a$ and $q_{z^{2}}$ ($----$ in both cases) in \Cro is consistent with its established anisotropic linear diagonal ME effect. On the other hand, in \Feo the antiferroic ordering of $a$ and $q_{z^{2}}$ ($++--$ in both cases) suggests an antiferroic linear diagonal ME response, in which an external electric field induces magnetic moments parallel to the electric field, but in opposite directions on Fe$_1$ and Fe$_2$ relative to Fe$_3$ and Fe$_4$, such that an antiferromagnetic order with no \emph{net} magnetic moment is established along the field direction. We refer to this response as an \textit{anti}-ME effect.
Furthermore, the antiferroically-ordered $t_z$ in both materials, two orders of magnitude smaller than $a$ and $q_{z^{2}}$, are indicative of an additional off-diagonal anti-ME effect, with the induced magnetic moments ordered differently in \Cro (Cr$_1$ and Cr$_4$ having opposite sign  relative to Cr$_2$ and Cr$_3$) and \Feo (Fe$_1$ and Fe$_3$ having opposite sign relative to Fe$_2$ and Fe$_4$).

We note that, in both materials, local ME multipoles are allowed on the oxygen sites as well, as reported in Tab. \ref{tab:Multi_CrFe_O_sign_dependent}, where the absolute signs are obtained from the first-principles calculations.
The Wyckoff site symmetry (2) of the O atoms does not include the three-fold axis, thus allows multipoles with non-zero in-plane components ($t_x$, $t_y$, $q_{xz}$, $q_{yz}$, $q_{xy}$, and $q_{x^{2}-y^{2}}$), in addition to the $a$ and $q_{z^2}$ also found on the TM ions, while it prohibits $t_z$. Out of all the multipole components on the O atoms, the only ones ordered ferroically are $a$ and $q_{z^{2}}$ in \Crop, indicating that the O atoms also contribute to the net ME effect in this material. All the other components sum up to zero, as dictated by the global symmetry, hence they do not contribute to a net ME effect, but rather to additional anti-ME responses.
\begin{table}[t]
    \centering
    {\renewcommand{\arraystretch}{1.5}
    \begin{tabular}{l|c|c|c|c|c|c}
        \hline
         & O$_1 $ & O$_2$ & O$_3$  & O$_4$ & O$_5$ & O$_6$ \\ 
        \cline{2-7} 
        $m_x$  & $m$ & $-\frac{1}{2}m$ & $-\frac{1}{2} m$ & $\mp m$  & $\pm \frac{1}{2}m$ & $\pm \frac{1}{2}m$ \\
        $m_y$  &  0    & $\frac{\sqrt{3}}{2} m$ & $-\frac{\sqrt{3}}{2} m $ &  0  & $\mp \frac{\sqrt{3}}{2} m$ & $\pm \frac{\sqrt{3}}{2} m$ \\
        \cline{2-7} 
        $a$  & $\mp a$ & $\mp a$ & $\mp a$ &  $-a$ & $-a$ & $-a$\\
        $t_x$ & $+t $& $ -\frac{1}{2}t $ &  $-\frac{1}{2}t$ &  $\pm t $& $\mp \frac{1}{2}t$ &  $\mp \frac{1}{2}t$\\
        $t_y$ & $ 0  $& $+\frac{\sqrt{3}}{2}t $ &  $-\frac{\sqrt{3}}{2}t$ &  $ 0  $& $ \pm \frac{\sqrt{3}}{2}t $ &  $ \mp \frac{\sqrt{3}}{2}t$ \\
        $q_{xy}$ & $ 0$& $\mp \frac{\sqrt{3}}{2}q_{2} $ &  $\pm \frac{\sqrt{3}}{2}q_{2}$ &  $ 0$& $ - \frac{\sqrt{3}}{2}q_{2} $ &  $ +\frac{\sqrt{3}}{2}q_{2}$ \\
        $q_{xz}$ & $ 0$& $+\frac{\sqrt{3}}{2}q_{1} $ &  $-\frac{\sqrt{3}}{2}q_{1}$ &  $ 0$& $\pm \frac{\sqrt{3}}{2}q_{1} $ &  $\mp \frac{\sqrt{3}}{2}q_{1}$ \\
        $q_{yz}$ & $-q_{1} $& $+\frac{1}{2}q_{1} $ &  $+\frac{1}{2}q_{1}$ &  $\mp q_{1} $& $\pm \frac{1}{2}q_{1} $ &  $ \pm \frac{1}{2}q_{1}$\\
        $q_{x^{2}-y^{2}}$ & $ \pm q_{2} $& $ \mp \frac{1}{2}q_{2} $ &  $\mp \frac{1}{2}q_{2}$ &  $ +q_{2} $& $ -\frac{1}{2}q_{2}$ &  $- \frac{1}{2}q_{2}$\\
        $q_{z^{2}}$ & $ \mp q_{3} $& $ \mp q_{3}$ &  $\mp q_{3}$&  $-q_{3}$ & $-q_{3}$&  $-q_{3}$ \\
        \hline
    \end{tabular}     
    }
    \caption{Symmetry-allowed magnetic moments and ME multipoles, and their ordering on the O atoms in \Cro and \Feop. When the sign is different in the two materials, two signs are given, with the top sign corresponding to \Cro and the bottom one to \Feop, respectively. The magnitudes (in $\mu_{\text{B}}$) of $m$ and the angular parts of $a$, $t$, $q_1$, $q_2$, $q_3$ are $7\times10^{-5}$, $1\times10^{-2}$, $1\times10^{-2}$, $2\times10^{-2}$, $4\times10^{-5}$, $1\times10^{-2}$ in \Cro and $2\times10^{-3}$, $3\times10^{-2}$, $2\times10^{-2}$, $2\times10^{-2}$, $3\times10^{-4}$, $4\times10^{-2}$ in \Feo respectively. Atoms are labelled as in Fig. \ref{fig:Mag_compare}.}
    \label{tab:Multi_CrFe_O_sign_dependent}
\end{table}

In addition to the ME multipoles associated with the linear ME effect, magnetic octupoles are also symmetry allowed. The relevant non-zero components on the TM ions are $\mathcal{O}_{-3}$ and $\mathcal{O}_{3}$, following the naming convention of Ref. \cite{urru_magnetic_2022} and, for an applied electric field along $y$, are associated with a local quadratic response in $m_y$ and $m_x$, respectively \cite{urru_magnetic_2022, rotation_footnote}. $\mathcal{O}_{3}$ and $\mathcal{O}_{-3}$ are ordered antiferroically ($--++$ and $-+-+$, respectively) in \Crop, and correspond to a second-order diagonal and off-diagonal anti-ME effect \cite{urru_magnetic_2022}. In \Feop, $\mathcal{O}_{-3}$ orders antiferroically ($-++-$) as well, but $\mathcal{O}_{3}$ orders ferroically ($++++$), which suggests that the lowest order \emph{net} ME response is the second-order off-diagonal ME effect.

%\subsection{\label{sec:me_response} Lattice-mediated local ME response}

\begin{figure}[t]
 \centering
  \includegraphics[width=0.48\textwidth]{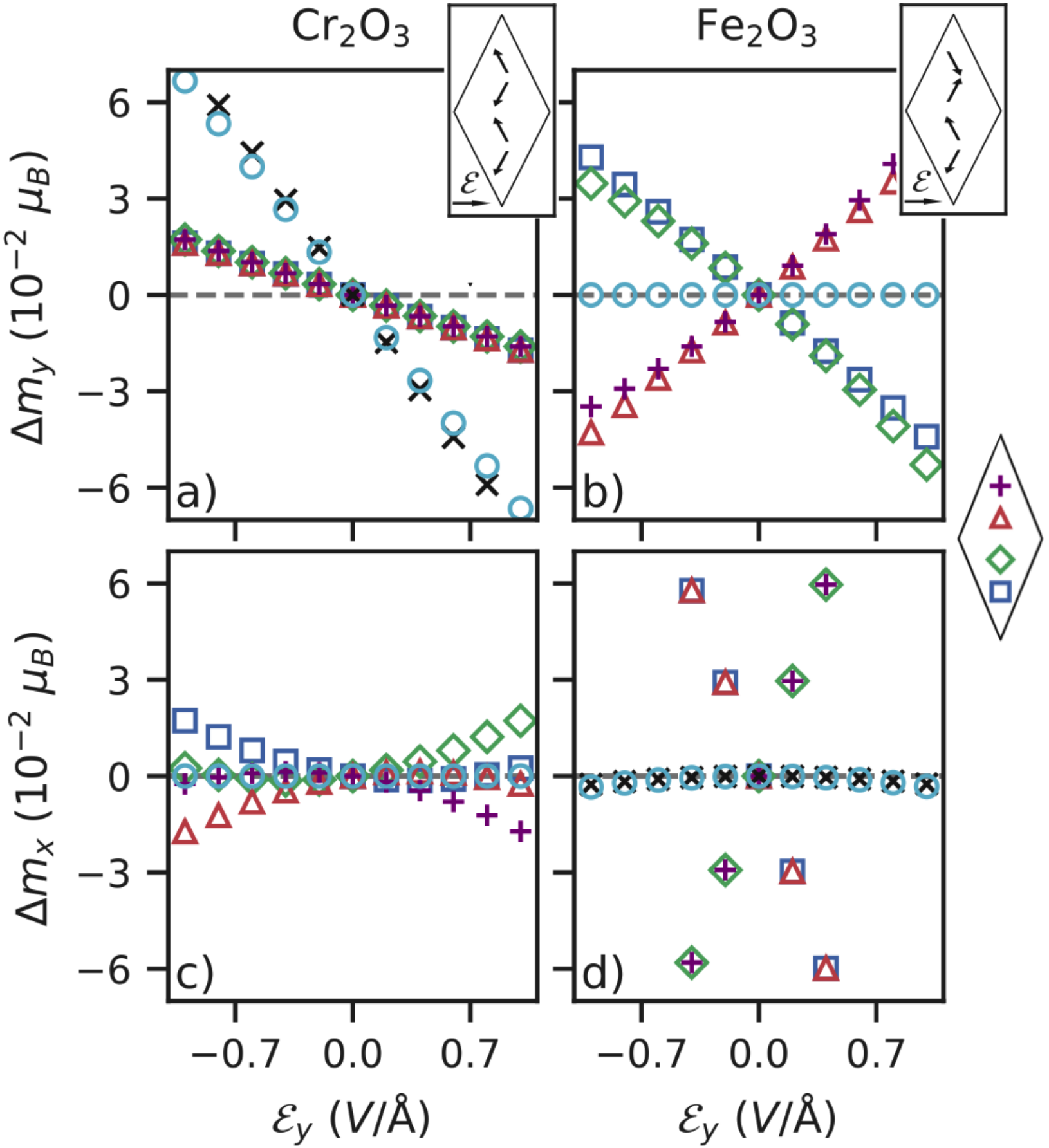}
\caption{Local change in the in-plane magnetic moments ($\Delta m_i$) on the TM ions as a function of the applied electric field strength, with $\Delta m_i$ parallel (a,b) and perpendicular (c,d) to the applied electric field direction, in \Cro (a,c) and \Feo (b,d). Blue squares, green diamonds, red triangles, and purple crosses represent the local $\Delta m_i$ on TM ions 1-4, respectively (see inset). The sum of the induced magnetic moments on the four TM ions (cyan circles), and the total induced magnetic moment in the unit cell (black diagonal crosses in a,d) are shown as well. Insets in a and b sketch the (exaggerated) linear response parallel to the applied $\mathcal{E}$, by showing the induced magnetic moments on top of the equilibrium magnetic order.}
\label{fig:ME_response}
\end{figure}

Now we use \textit{ab initio} density functional theory to calculate the anti-ME effects predicted by the symmetry arguments discussed above, in \Cro and \Feop. Fig. \ref{fig:ME_response}, which summarizes our results, shows the calculated lattice-mediated changes in the magnetic moments induced by an electric field pointing along the positive $y$ direction, for both \Cro (panels a and c) and \Feop (panels b and d), for the antiferromagnetic domains shown in Fig. \ref{fig:Mag_compare}. We separately consider the induced moments along $y$ (panels a and b), associated with a diagonal ME response, and along $x$ (panels c and d), associated with an in-plane off-diagonal response \cite{out_of_plane_footnote}. 
In Fig. \ref{fig:ME_response}a, we see that the moments on all the four Cr atoms in \Cro show an identical linear dependence on the strength of the applied electric field. This indicates an identical local diagonal linear ME response, adding up to a net diagonal linear ME effect over the unit cell. This is consistent with the ferroic ordering of $a$ and $q_{z^{2}}$ on the Cr ions. Furthermore, the sum of the induced local magnetic moments on the Cr atoms (cyan circles) is almost identical to the total induced magnetic moment per unit cell (black diagonal crosses), showing that the response is dominated by the Cr atoms.    
%total response over the unit cell is a bit larger than the summed response of the Cr ions, indicating a contribution from the O ions, as predicted from the ferroic ordering of $a$ and $q_{z^{2}}$ on those. 
We remark that the sign of the response matches that found in previous first-principles calculations \cite{malashevichFullMagnetoelectricResponse2012, iniguezFirstPrinciplesApproachLatticeMediated2008}. % and experiment \cite{brownDeterminationMagnetizationDistribution2002}. 
Although not visible in the plot, there is an additional small quadratic component to the induced magnetic moments as a function of electric field strength, but summed over the atoms this cancels out. This diagonal second-order anti-{ME} effect is consistent with the antiferroic ordering of the ${\mathcal{O}}_{-3}$ octupoles mentioned above.

The local induced magnetic moments parallel to the applied electric field on the four Fe atoms in \Feo (Fig. \ref{fig:ME_response}) show linear dependence on the field strength of identical magnitude, but opposite sign; the induced moments order pairwise, with opposite sign for the two pairs of Fe ions, resulting in no net induced magnetic moment in the unit cell.
This linear anti-ME effect, consistent with the antiferroic ordering of $a$ and $q_{z^{2}}$ discussed before, is the lowest order ME response in \Feo and, to the best of our knowledge, has not been previously discussed. In addition to the linear contribution, we note the presence of a small local quadratic response, consistent with the antiferroic ordering of the ${\mathcal{O}}_{-3}$ octupoles.

Next, we consider the induced in-plane magnetic moments perpendicular to the applied electric field, corresponding to the off-diagonal in-plane ME response. In \Cro (Fig. \ref{fig:ME_response}c), these moments show a linear as well as a quadratic dependence on the strength of the applied electric field, but both contributions cancel out to make the net response zero. This indicates both a linear and quadratic off-diagonal anti-ME effect, which is expected from the antiferroic order of  $t_z$ ($+--+$) and the $\mathcal{O}_{3}$ octupoles ($--++$). 

Finally, in \Feo (Fig. \ref{fig:ME_response}d), the induced in-plane magnetic moments perpendicular to the applied electric field have a dominant anti-aligned linear dependence, with opposite sign on different pairs of Fe atoms. The linear part of the induced moments sums to zero, leading to no net induced moment in the unit cell. This corresponds to an off-diagonal anti-ME effect, following from the antiferroic ordering of $t_z$, similarly to \Crop. We remark that in both \Cro and \Feo the linear diagonal and off-diagonal responses are of the same order of magnitude. 
Interestingly, there is also a smaller quadratic dependence. As the summed (cyan circles) and total (black diagonal crosses) induced moments reveal, this contribution is ferroic and does not sum to zero, instead indicating a net bulk second-order ME response. This is thus the lowest order ferroic ME response in \Feop, and follows from the ferroic ordering ($++++$) of the $\mathcal{O}_{3}$ octupoles.  

In Tab. \ref{tab:conclusion} we summarize the ME responses discussed above. We note that the proposed anti-ME effect is more ubiquitous than the ferroic ME effect, since it follows from less restrictive symmetry requirements. As a consequence, a substantial fraction of magnetic materials is expected to show a local anti-ferroically ordered ME response.

\begin{table}[t]
    \centering
    {\renewcommand{\arraystretch}{1.5}
    \begin{tabular}{c|c c|c|c|c}
        \hline
              & Diagonal & Off-diagonal \\
        \hline
        \Cro  & L : ferro ME  &  L : anti-ME \\
              & Q : anti-ME  &  Q : anti-ME \\
        \hline
        \Feo  & L : anti-ME   &  L : anti-ME  \\
              & Q : anti-ME  &  Q : ferro ME  \\
        \hline
        \end{tabular}     
         }
    \caption{Summary of the in-plane ME effects found in \Cro and \Feop. L and Q indicate linear and quadratic ME effects, respectively.}
    \label{tab:conclusion}
\end{table}

%\section{\label{sec:sumout} Summary \lowercase{and} Outlook}
In this work, we studied the connection between the local ME multipolar order and the local atomic ME response. We discussed as case studies the prototypical ME material \Cro and the centrosymmetric material \Feop. Beyond the well established linear diagonal ME in \Crop, we predicted via symmetry and multipole analysis an off-diagonal anti-ME in \Cro as well as both diagonal and off-diagonal anti-ME effects in \Feop, and confirmed our predictions using \textit{ab initio} calculations. Additionally, we found in both materials a non-negligible local second-order ME response, which sums to a net response in \Feop, and which we rationalized with the presence of magnetic octupoles. 

Our findings allow us thus to broaden the concept of ME response in ordered materials: to have a local ME response, no global symmetry breaking is required, hence even materials that preserve both inversion and time-reversal (the latter with a fractional translation), e.g. NiO, allow for a non-zero local ME tensor. The only strict requirement to have any local ME effect is the breaking of time-reversal at the Wyckoff site. This means that materials belonging to magnetic space groups (MSGs) of type I (colorless), III or IV (black-white) allow local ME response; ordered materials of MSG II (grey), instead, do not show any local ME response. If, besides local time-reversal breaking, at least one atomic species sits in a Wyckoff site that is not an inversion center, e.g. Mn$_3$O$_4$ \cite{yuan_prediction_2021}, a local \emph{linear} ME response is allowed, otherwise the lowest order response is quadratic.

Our calculations show that the local linear anti-ME response is of the same order of magnitude as the local ferro-ME response in similar non-centrosymmetric materials. Thus, the main challenge in measuring an anti-ME response is not the size of the response, but rather the anti-alignment of the induced magnetic moments, producing a vanishing net ME response. 

In order to measure and possibly exploit such an anti-ME response, an external electric field varying at the length scale of the unit cell would be desirable as it would induce a net magnetization; this could alternatively be achieved by exciting a coherent phonon with the appropriate pattern of polar atomic displacements. We hope that our findings motivate further experimental investigations to measure such anti-ME effects, as well as theoretical studies to identify promising candidates with effects of larger size.

\begin{acknowledgments}
The authors thank Dr. Michael Fechner, Dr. John Kay Dewhurst and Dr. Sayantika Bhowal for useful discussions. NAS, XHV, and AU were supported by the ERC under the European Union’s Horizon 2020 research and innovation programme grant No. 810451 and by the ETH Z\"urich. Computational resources were provided by ETH Z\"urich’s Euler cluster.
\end{acknowledgments}

%\appendix

\bibliographystyle{apsrev4-2}
\bibliography{apssamp,references,references_2}% Produces the bibliography via BibTeX.

\begin{thebibliography}{42}
\bibitem [{lat()}]{lattice_parameter_footnote}%
  \BibitemOpen
  \href@noop {} {\bibinfo {title} {Although our equilibrium lattice parameters
  are reduced with respect to the experimental ones, we still are still able to
  capture the same physics. {F}ixing the lattice parameters to the experimental
  values, did not qualitatively alter our results.}}\BibitemShut {Stop}%
\bibitem [{out()}]{out_of_plane_footnote}%
  \BibitemOpen
  \href@noop {} {\bibinfo {title} {The linear out-of-plane response to an
  in-plane applied electric field is zero, so we do not consider it
  here.}}\BibitemShut {Stop}%
\end{thebibliography}

\end{document}